\documentclass{article}

\usepackage[citestyle=numeric,bibstyle=numeric,maxcitenames=2,maxbibnames=9,uniquelist=false]{biblatex}
\usepackage[nonatbib]{neurips_2023}

\usepackage[utf8]{inputenc} 
\usepackage[T1]{fontenc}    
\usepackage{hyperref}       
\usepackage{url}            
\usepackage{booktabs}       
\usepackage{amsfonts}       
\usepackage{nicefrac}       
\usepackage{microtype}      
\usepackage{xcolor}         

\usepackage{amsmath}
\usepackage{amsthm}
\usepackage{placeins}
\usepackage{microtype}
\usepackage{graphicx}
\usepackage{subfigure}
\usepackage{booktabs} 
\usepackage{thmtools, thm-restate}
\DeclareMathOperator*{\argmax}{arg\,max}
\DeclareMathOperator*{\argmin}{arg\,min}
\usepackage{amsfonts}
\usepackage{enumitem}

\usepackage{cleveref}

\definecolor{bored}{HTML}{EF553B}
\definecolor{esmblue}{HTML}{636ef9}

\newcommand{\robot}{\texttt{ROBOT}} 
\newcommand{\bx}{\mathbf{x}}
\newcommand{\bz}{\mathbf{z}}

\newcommand{\by}{\mathbf{y}}

\newcommand{\calE}{\mathcal{E}}

\newcommand{\citet}[1]{\textcite{#1}}
\newcommand{\citep}[1]{\parencite{#1}}

\title{Inverse Protein Folding Using \\ Deep Bayesian Optimization}

\begin{document}

\maketitle

\begin{abstract}
Inverse protein folding---the task of predicting a protein sequence from its backbone atom coordinates---has surfaced as an important problem in the ``top down'', \emph{de novo} design of proteins. Contemporary approaches have cast this problem as a conditional generative modelling problem, where a large generative model over protein sequences is conditioned on the backbone. While these generative models very rapidly produce promising sequences, independent draws from generative models may fail to produce sequences that reliably fold to the correct backbone. Furthermore, it is challenging to adapt pure generative approaches to other settings, e.g., when constraints exist. In this paper, we cast the problem of improving generated inverse folds as an optimization problem that we solve using recent advances in ``deep'' or ``latent space'' Bayesian optimization. Our approach consistently produces protein sequences with greatly reduced structural error to the target backbone structure as measured by TM score and RMSD while using fewer computational resources. Additionally, we demonstrate other advantages of an optimization-based approach to the problem, such as the ability to handle constraints. 
\end{abstract}

\section{Introduction}
\label{sec:intro}
\emph{De novo} protein design, the design of amino acid sequences that will fold into protein structures that achieve a set of desired biochemical or functional properties, is one of the central bioengineering challenges of the 21st century \citep{denovo_proteins, denovoprot1, denovoprot2, denovoprot3}. Recent advances in computational methods for protein folding \citep{alphafold2,esmfold,alphafoldmultimer,rosettafold} have led to the rise in these ``top-down'' approaches to protein design, where one directly starts with protein structures that achieve some goal, and seek amino acid sequences that achieve that structure. This approach is especially promising when coupled with recent work like RFdiffusion \citep{rfdiff}, which \textit{directly generate} protein structures that achieve desired properties.

This approach to protein design requires a solution to the \textit{inverse folding problem} \citep{yue1992inverse,physics_based,jing2021equivariant} of designing amino acid sequences that fold into a given backbone structure. Recent inverse folding approaches using large language modelling have surged in accuracy given the new wealth of accurate, computationally-determined protein structure data available, and now achieve state of the art performance \citep{esmif, if_1, if_2, if_3, if_4, joint_seq_struct_1, joint_seq_struct_2}. 

However, these approaches have tangible disadvantages in practice. In protein engineering, it's common to devote significant resources to solving a specific, focused design problem, often under numerous developability and design constraints. Generative and one-shot prediction approaches, however, are more suitable for ``wide'' results, demonstrating the success of one-shot or few-shot attempts to inverse fold large databases of structures, rather than deep, focused efforts to inverse fold a handful of specific structures with high accuracy using iterated feedback. The focused setting, where one seeks to inverse fold specific target structures very well rather than achieve good performance on average, is arguably more important in protein engineering.

In this paper, rather than solving inverse folding through pure generation, we develop a blackbox optimization-focused pipeline that leverages recent rapid advancements in ``latent space'' or ``deep'' Bayesian optimization \cite{maus2022local, lambo, Gligorijevi2021, BOSS, Winter2019, SanchezLengeling2018, GmezBombarelli2018, Griffiths2020, grosnit2021highdimensional}, where generation provides initial solutions that are iteratively improved. Given an amino acid sequence $\bx$ and an objective function $f(\bx)$ that measures (for example) the similarity between the computationally folded structure of $\bx$ and the target structure, we seek to maximize $f(\bx)$. Concretely, we make the following contributions:

\begin{enumerate}[wide, labelwidth=!, labelindent=5pt, leftmargin=20pt]
    \item In contrast to recent work on inverse protein folding, we cast the task as an optimization problem. Rather than solving inverse folding as a one shot prediction task, this enables us to iteratively refine the design of a sequence, resulting in sequences that fold computationally to significantly better matches. Furthermore, our approach uses a significantly smaller generative model, enabling it to use less computational resources overall.
    
    \item We develop an optimization pipeline, \texttt{BO-IF}, leveraging recent work on latent space Bayesian optimization. Our approach deploys more accessible computational resources, using a ``small'' transformer language model with only 47M parameters, trainable in a few days on a single RTX A6000. We use this model in concert with Bayesian optimization to solve inverse protein folding problems, and make our pipeline publicly available using standard software libraries like BoTorch \cite{balandat2020botorch} at \url{https://github.com/nataliemaus/bo-if}.
    
    \item We demonstrate that our method has substantial advantages over pure generation:
    \begin{enumerate}
        \item \textbf{Accuracy.} Our approach \textbf{reduces the structural error} of sequences produced by ESM-IF \citep{esmif} to the target \textbf{by 48$\%$ on average} as measured by 1-TM score, and \textbf{28$\%$} as measured by RMSD.
        \item \textbf{Efficiency.} On a single GPU, the end-to-end optimization time considering and folding 150,000 sequences \textit{sequentially} is roughly equivalent to parallel generation and folding due to the size of the generative models used, with both approaches taking approximately 50 GPU hours.
    \end{enumerate}

    \item We extend our method to additional settings, using extensions of Bayesian optimization (BO) from the literature. In particular:
    \begin{enumerate}
        \item We demonstrate our approach's ability to optimize under blackbox constraints by adapting constrained BO to this setting.
        \item We demonstrate that we are able to design \textit{diverse sets} of sequences that fold to the target structure by leveraging recent work on using BO for diverse generation. \citep{maus2022discovering}.
    \end{enumerate}
\end{enumerate}
\vspace{-2ex}
\begin{figure}[ht]
\begin{center}
\includegraphics[width=0.9\textwidth]{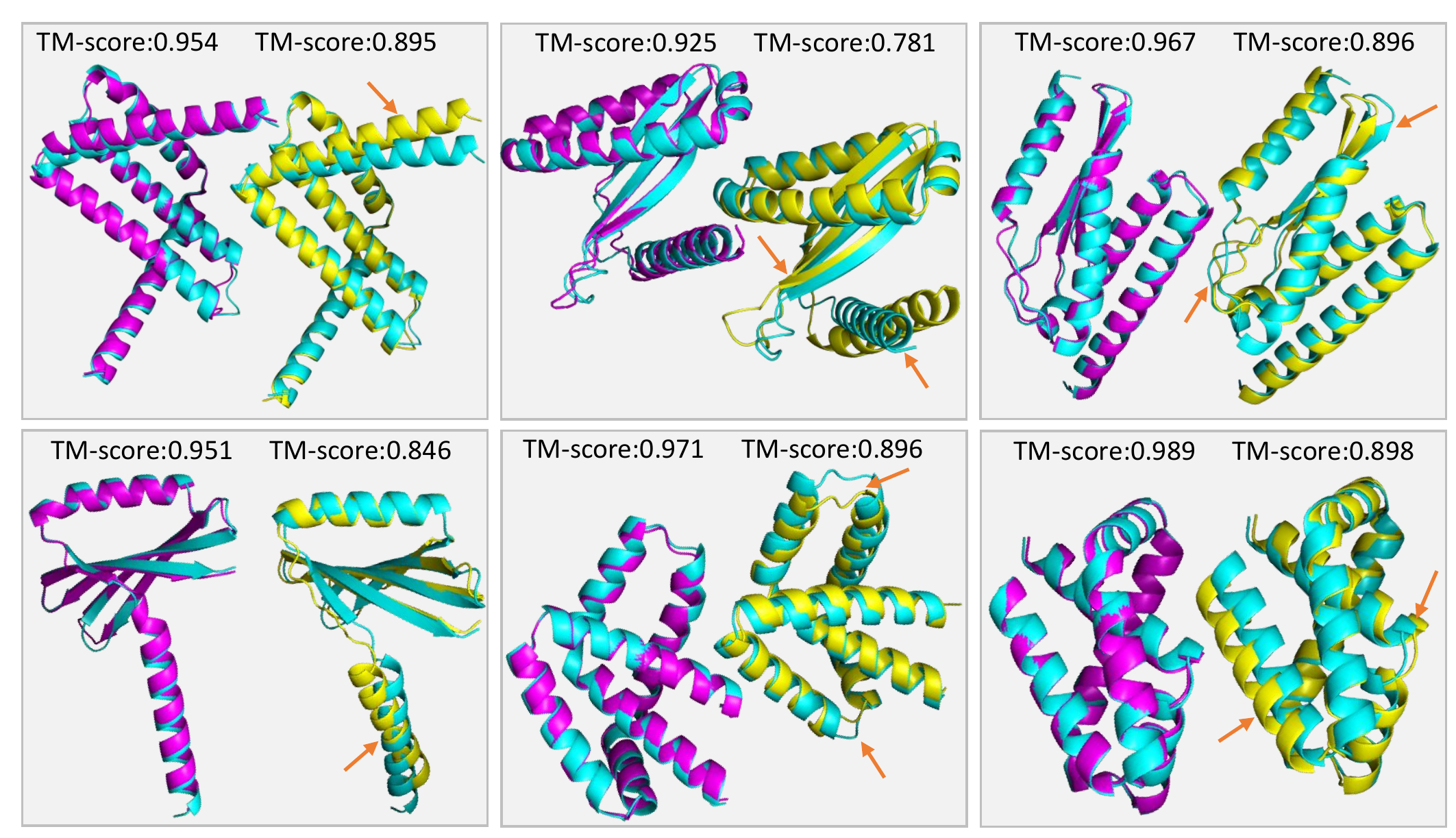} 
\caption{Different target backbones (blue) inverse folded by ESM-IF (yellow) and Bayesian optimization (pink). Our method, \texttt{BO-IF}, consistently finds proteins that better match the target structure as evidenced visually by the better alignment and by the higher TM-scores achieved. Arrows indicate example regions of mismatch.} 
\label{sample25}
\end{center}
\vskip -0.3in
\end{figure}

\section{Background and Related Work}
\label{sec:background}

\textbf{Computational protein folding.} Utilizing a Convolutional Neural Network (CNN) backbone and starting from features derived from multiple sequence alignment (MSAs), AlphaFold demonstrated the feasibility of learning a protein-specific potential and predicting a protein's structure from its sequence through potential minimization via gradient descent \cite{alphafold1}. However, analogous to earlier approaches, its performance declines in the absence of homologous structures \citep{homologous_prot, alphafold2}. AlphaFold2 further improved upon this by incorporating an SE(3)-equivariant Transformer network for refining atomic coordinates, thereby enhancing generalization capabilities to structures without homologs \citep{alphafold2}. Subsequent advancements, including RosettaFold and AlphaFold Multimer, have further refined these methodologies, enabling the generation of accurate models of protein-protein complexes \citep{alphafoldmultimer, rosettafold}. ESM-Fold tackled this problem from a sequence-only perspective, building on a 15-billion parameter protein language model \cite{esmfold, esm_1}, achieving near parity with alignment-based methods while being orders of magnitude faster. This significant acceleration in folding is what facilitates our exploration of the inverse folding problem through optimization and sampling in our current work.

\textbf{Designing sequences for target functionality.} Across the many subdomains of bioengineering, the three dimensional structure and topology of proteins determine their function \cite{Chiu2019-na, biomaterials-Varanko2020-fe, diagnostics-Brennan2010-wc}.
The field of \emph{de novo} protein design attempts to generate new amino acid sequences---with no necessary relationship to those found in nature---that achieve a desired complex fold, and thus produce desired function. Over the last several decades, many approaches have been developed to solve the task of computational protein design, with most focus on statistical approaches \cite{statistics-Sippl1990-gq, statistics2-Schaeffer2011-id} and expensive biophysical simulations \cite{physics_based, physics2-Brini2020-wm}. Only recently have deep learning methods been shown to take advantage of the wealth of known \cite{PDB} or predicted \cite{alphafold2} protein structures. 
\par

\cite{korteme} describe ``protein design as an optimization problem: given a user-defined structure and function, find one or a few low-energy amino acid sequences stably adopting the desired structure and performing the targeted function.'' Some of the largest barriers to computational protein design have only been addressed recently. For example, generating a reliable protein backbone \emph{de novo} has been made more accessible by RFDiffusion \cite{rfdiff}. 
Nonetheless, it is remains difficult to confidently describe the sequence or set of sequences that may be optimal for a specified structure\cite{Yang2019}, to ensure those sequences meet necessary design constraints \cite{Huang2016}, and to fully validate the function of designed sequences with meaningful functional readouts \cite{Tinberg2013}.

While existing protein design methods are able to output unconstrained amino acid sequences that fold into a specific structure \cite{rfdiff, korteme, esmif}, real-world protein design applications require that those proposed sequences satisfy both the inverse-folding challenge \textit{and} conform to arbitrary constraints. 
These additional constraints enable the production and characterization of those proteins or ensure their safe and effective use in research, industrial, diagnostic or therapeutic settings. 
Specifically, a high-throughput protein engineering platform could process tens to hundreds or more candidate sequences in a single batch \cite{ht-protein-expr-Doyle_undated-bf, ht-protein-expr2-Vincentelli2011-ah}. 
However, the designed proteins would require specific sequence constraints to ensure sufficient protein yield, stability and solubility \cite{sasa-Ali2014-gk, electrostatics-Vascon2020-be}. 
Alternatively, for downstream applications like therapeutic development, constraints would be needed to avoid unnecessary immunogenicity via humanization \cite{PrihodaHumanization} and/or minimization of aggregation propensity \cite{viscosity-Rai2023-by}.

\textbf{Black-box and Bayesian optimization.} 
In black-box optimization, we seek to find the minimizer of a function, $\argmin_{\mathbf{x}} f(\mathbf{x})$.
Commonly, $f(\mathbf{x})$ is assumed to be expensive to evaluate and unknown (i.e., a ``black box''). Classically, this problem has been considered in the setting where the search space is continuous. However, many applications across the natural sciences and engineering require optimizing over discrete and structured input spaces, such as amino acid sequences.

Bayesian optimization (BO) \citep{osborne2009gaussian, movckus1975bayesian, SnoekBO} is an approach to sample-efficient black-box optimization. At iteration $t$ of Bayesian optimization, one has access to a dataset $\mathcal{D} = [(\mathbf{x}_{i}, y_{i})]_{i=1}^{t}$, where $y_{i}$ denotes the (possibly noisy) objective value of the input $\bx_{i}$. This data is used to build a probabilistic \textit{surrogate model}---commonly a Gaussian process (GP) \citep{rasmussen2003gaussian}---of the objective function. This supervised surrogate informs a policy---often called an \textit{acquisition function}---about what sample to evaluate next. After the objective function is evaluated on a candidate, this observation is added to the dataset $\mathcal{D}$ and the surrogate model is updated in iteration $t + 1$. BO then proceeds by collecting observations of the objective function in an iterated sequential fashion, successively building a larger dataset $\mathcal{D}$ and therefore a better supervised model of the objective which results in a better policy. 

\textbf{Bayesian optimization for biological discovery.} Bayesian Optimization (BO) has become a powerful tool for biological discovery due to the vast space of potential biological sequences and structures. Several methods have been proposed that apply BO over discrete spaces \citep{moss2020boss, bocs}. Notably, a stream of research has used BO to discover new molecules within the latent spaces of Variational Autoencoders \citep{kingma2013auto, gomez2018automatic}. These methods either directly encode and decode from string representations \citep{maus2022local, lambo}, or utilize more complex graphical or grammatical structures \citep{JTVAE, kusner2017grammar}. Concurrently, BO has been applied to optimize protein sequences for desired functionalities. Techniques such as \citet{antbo} and \citet{romerofitness2013} have explored the protein fitness landscape and designed new antibodies using BO, while \citet{lambo} explored Bayesian optimization for sequence design without the need for a large pretraining corpus. These advances demonstrate the potential and effectiveness of BO in the field of biological discovery. 
\section{Inverse Folding as Optimization}
\label{sec:methods}
Inverse folding seeks to design amino acid sequences that fold to a desired structure. Commonly, one focuses on backbone structure alone, ignoring side chains \cite{esmif}. Formally, we are given as input a sequence of spatial coordinates $\bx = (x_{1}, ..., x_{n})$, with $x_{i} \in \mathbb{R}^{3k}$ representing the 3D spatial coordinates of $k$ backbone atoms per amino acid in the structure. The goal is to produce a sequence of amino acids $\by = (y_{1},...,y_{n})$ that folds to a structure with backbone coordinates determined by $\bx$.

\subsection{Bayesian Optimization Strategy}

\textbf{Inverse folding as generation.} In a generative approach to inverse folding, \citet{esmif} train an auto-regressive model of the conditional distribution,
\begin{equation*}
 \Psi(\by \mid \bx) = \prod_{i=1}^{n} p(y_{i} \mid y_{1:i-1}, \bx),
\end{equation*}
where a supervised dataset of structure and native sequence pairs $[(\bx_{i},\by_{i})]_{i=1}^n$ is collected using a combination of experimentally determined structures and high confidence structures produced using computational folding techniques \citep{alphafold2,esmfold}. \citet{esmif} train $\Psi$ as a GVP-GNN \citep{jing2021equivariant} followed by a generic encoder-decoder Transformer \citep{vaswani2017attention}. Given a new target sequence $\bx^{+}$, amino acid sequences may be sampled from the trained autoregressive model above, $\by^{+} \sim \Psi(\by \mid \bx^{+})$.

\textbf{Inverse folding as optimization.} To formulate the inverse folding task as an optimization problem, we formulate an objective function that measures how closely the sequence $\by$ folds to the structure $\bx^{+}$ and then optimize it. Letting $\mathcal{F} : \mathcal{Y} \to \mathcal{X}$ denote a transformation that maps sequences to backbone atom coordinates using a computational folding model like AlphaFold2 or ESMFold, we seek to solve optimization problems of the form
\begin{equation}
    \by^{+} = \argmin_{\by} \mathcal{E}(\mathcal{F}(\by), \bx^{+}), \label{eq:opt_problem}
\end{equation}
where $\calE(\bx, \bx')$ measures the structural error between $\bx$ and $\bx'$. For example, $\calE$ could be taken to be $1-$TM score \citep{tm_score}, or RMSD. We first focus on solving the unconstrained optimization problem as presented above before discussing extensions including constraints and finding diverse solutions.

Even with access to computational folding models, this optimization problem is challenging because it is over the discrete space of $20^n$ possible amino acid sequences of length $n$. In the experiments we conduct, $n$ ranges from 100 to 150. Furthermore, fast computational folding models like ESMFold \cite{esmfold} are sufficiently large that folding a sequence takes on the order of seconds on average, even when batched to the maximum capacity of an RTX A6000. We are therefore moderately computationally limited, and exploring a few hundred thousand sequences takes on the order of a few GPU days.

\textbf{Latent space Bayesian optimization} To solve the optimization problem, \autoref{eq:opt_problem}, we utilize recently developed latent space Bayesian optimization techniques that adapt Bayesian optimization from continuous blackbox optimization problems to discrete and structured ones \citep{Weighted_Retraining,ladder,gomez2018automatic,Huawei,eissman2018bayesian,kajino2019molecular,maus2022local,lambo}. Latent space Bayesian optimization seeks to leverage the representation learning capabilities of deep generative models, most commonly variational autoencoders (VAEs) \cite{kingma2013auto} to aid in optimization. 

Briefly, a VAE consists of an encoder $\Phi(\bz \mid \by) : \mathcal{Y} \to \mathcal{P}(\mathcal{Z})$ mapping from amino acid sequences $\mathcal{Y}$ to a distribution over a continuous latent space $\mathcal{Z}$, and a decoder $\Gamma(\by \mid \bz) : \mathcal{Z} \to \mathcal{P}(\mathcal{Y})$ that (probabilistically) reverses this process. At a high level, the idea is to perform optimization over the latent space $\mathcal{Z}$ of the VAE, rather than the discrete and structured space of amino acid sequences $\mathcal{Y}$. This constrains the optimizer to sequences in $\mathcal{Y}$ that the VAE can generate, but simplifies the optimization problem considerably in return. Given a trained VAE, the optimization problem in \autoref{eq:opt_problem} becomes continuous:
\begin{equation*}
   \by^{+} \approx \Gamma(\bz^{+})
   \quad\text{where}\quad
   \bz^{+} = \argmin_{\bz \in \mathcal{Z}} \mathcal{E}(\mathcal{F}(\Gamma(\bz)), \bx^{+}) \label{eq:latent_opt_problem}
\end{equation*}
Here, we abuse notation and use $\Gamma(\bz):=\argmax_{\by}\Gamma(\by\mid\bz)$ to denote the most likely decoding of the latent vector $\bz$---note that in this setting, we are generally uninterested in the expected behavior of the objective over the \textit{distribution} $\Gamma(\by \mid \bz)$, because we are ultimately interested in the final sequence $\by^{+}$. This objective function takes a latent code $\mathbf{z}$, generates a sequence $\by$ via the decoder, which is then folded by $\mathcal{F}$ and evaluated. Standard Bayesian optimization can now be directly applied to the maximization over $\bz$. We use \texttt{LOL-BO} \citep{maus2022local} as our base Bayesian optimization routine, although with larger VAE and surrogate models than used in the original small molecule setting.

\subsection{Model Architectures}

\textbf{Transformer VAE.} In this work, we pretrain an autoregressive VAE encoder-decoder Transformer architecture \cite{vaswani2017attention} with 6 encoder layers and 6 decoder layers totaling 47 million parameters on a randomly selected subset of 1.5 million protein sequences with lengths of 100-300 amino acids from Uniref \cite{uniref}. The encoder maps from amino acid sequences down to a total latent dimensionality of 1024, each amino acid is a separate token. The model was trained using the standard VAE ELBO with the KL divergence term multiplied by a factor of $10^{-4}$. 

\textbf{Surrogate model.} In order to support Bayesian optimization with hundreds of thousands of queries, we use sparse variational Gaussian process approximations \cite{hensman2013gaussian}, and in particular the parametric Gaussian process regressor (PPGPR) model of \cite{PPGPR}. This surrogate model is trained on pairs of latent codes $\bz$ and corresponding quality observations $\calE(\mathcal{F}(\Gamma(\bz)), \bx^{+})$. Because of the high dimensionality of the latent space and the sometimes poor performance of kernel methods on high dimensional data, we use a small deep kernel \cite{wilson2016deep} with two fully connected hidden layers of dimensionality 256 to reduce down to 256 dimensions. 

\subsection{Extensions}
\label{sec:extensions}

\textbf{Constrainted optimization.} Protein engineering is almost always done under numerous constraints that seek to limit cost, immunogenicity, and other risk factors. Extending \autoref{eq:opt_problem} to handle one or more constraints enables searching for amino acid sequence solutions that achieve desired design properties, for example high likelihood under a language model of ``natural'' proteins or high similarity to human-like proteins. We consider optimizing subject to constraints on the output sequence:
\begin{equation*}
    \bz^{+} = \argmin_{\bz \in \mathcal{Z}} \calE(\mathcal{F}(\Gamma(\bz), \bx^{+})) \; \textrm{s.t.} \; \forall i \, c_{i}(\Gamma(\bz)) \leq 0,
\end{equation*}
where $c_{i}(\cdot)$ are black-box constraints that might operate directly on the sequence $\by := \Gamma(\bz)$ or even on the computationally determined structure $\mathcal{F}(\by)$. In constrained Bayesian optimization (e.g., \citep{cei,gelbart2014bayesian,pesc,scbo}), additional surrogate models are trained to model the constraint functions $c_{i}(\cdot)$, and information from these additional surrogates is incorporated into the acquisition process. The most straightforward adaptation is to use \texttt{SCBO} \citep{scbo} as the optimization routine, which is the constrained analog of the \texttt{TuRBO} algorithm used by \texttt{LOL-BO} \cite{maus2022local}.

\textbf{Joint training of constraint surrogates.} \citet{maus2022local} found that end-to-end joint variational training of the VAE and GP surrogate model significantly improves optimization performance. Adapting this idea to the constrained setting, we train the constraint surrogate models end-to-end jointly with the objective surrogate and VAE. This involves optimizing the following ELBO, derived for the joint model over the $k$ GPs involved (1 for the objective, $k - 1$ for each of the $k - 1$ constraints):
\begin{equation*}
    \mathcal{L}_{\textrm{joint}}(\theta_{\Phi}, \theta_{\Gamma}, \theta_{\textrm{GP}_{1:k}}) = \mathbb{E}_{\Phi(\bz \mid \by) }\left[\sum_{i=1}^{k} \mathcal{L}_{\textrm{GP}_{i}}\left(\theta_{\textrm{GP}_{i}}, \theta_{\Phi} ; \bz, \mathbf{q}\right)\right] + \mathcal{L}_{\textrm{VAE}}(\theta_{\Phi}, \theta_{\Gamma} ; \by),
\end{equation*}
where $\mathbf{q}$ denotes the set of acquired structure errors obtained so far during optimization.

\textbf{Finding diverse solutions.} A common need in protein engineering is to diversify sequence solutions to diversify risk of a given design. This is particularly true in therapeutic development where each sequence carries unique liabilities into drug development. \texttt{ROBOT}~\citep{maus2022discovering} is an extension of the BO framework to enforce sequence diversity during optimization. We use edit distance $\delta(\by, \by')$ as a notion of diversity between two sequences, and solve the following series of optimization problems:
\begin{align*}
    \by^{+}_{1} &= \argmin_{\by} \calE(\mathcal{F}(\by), \bx^{+}) \\
    \by^{+}_{i} &= \argmin_{\by} \calE(\mathcal{F}(\by), \bx^{+}) \; \textrm{s.t.} \; \delta(\by, \by^{+}_{j}) \geq \tau \textrm{ for $j = 1,...,i - 1$}
\end{align*}
This procedure produces a set of low error sequences $\by^{+},...,\by^{+}_{i}$ so that each pair of sequences is separated by an edit distance of at least $\tau$.

\section{Experiments}
\label{sec:experiments}

We separate the evaluation of our method, which we call \texttt{BO-IF}, into two phases. First, we evaluate the optimization performance of applying our pipeline to backbone structures. Second, we evaluate extensions of our approach to the constrained setting and to finding diverse sets of high scoring sequences. We additionally evaluate the relative computational costs of both approaches.  

\textbf{Implementation details and hyperparameters.} We implement our pipeline leveraging BoTorch~\citep{balandat2020botorch} and GPyTorch~\citep{gardner2018gpytorch}, with code available at \url{https://github.com/nataliemaus/bo-if}. Other than the model architecture details specifically described previously in the paper, all hyperparameters for all Bayesian optimization methods used are set to the defaults used by their respective papers. Optimization runs is initialized with 1,000 sequences sampled from the ESM-IF model.

\textbf{Baseline.} Throughout this section, we use the GVP Transformer model of \citet{esmif} (also known as ESM-IF) as a recent, ``gold standard'' approach that (1) still achieves state-of-the-art performance for inverse folding via generation, and (2) maintains publicly available open source software. We initially evaluated performance with both the high and low temperature settings recommended by the authors ($T = 1e-6$ and $T = 1.0$). We found that repeated sampling with the lower temperature failed to improve TM score beyond the first handful of samples. We therefore report results against the higher temperature value of 1.0. We emphasize that repeated sampling is reasonably standard usage of these models -- e.g., this is the recommended approach of \citet{rfdiff} for finding an inverse fold to generated structures.

\subsection{Model Statistics and Efficiency}
We report statistics on pretraining and optimization runtime for both our approach and ESM-IF in \autoref{tab:vae_training}. For an input protein backbone structure during inference, both optimization and sampling were performed using a single RTX A6000 with 48 GB of VRAM. In order to fully utilize GPU resources, both approaches are capable of batch evaluation. 
We sample 20 samples from the ESM-IF decoder in batch parallel, and for Bayesian optimization, we leverage the batch acquisition capabilities of Thompson sampling to evaluate batches of 10 samples in parallel. 

These results suggest that iterative batch optimization using BO and decoding from a large language model have roughly comparable overhead for generating candidate sequences. Finally, we observe that roughly $83\%$ of the total running time for both methods is spent computationally folding candidate sequences using ESMFold~\citep{esmfold}. 

\subsection{Unconditional Inverse Folding of Protein Backbones}
Next, we evaluate whether using optimization to make focused, concerted efforts to inverse fold structures leads to improved structural similarity. Ideally, this evaluation should be done on roughly random protein structures that are unseen by both ESM-IF and our pretrained model. To accomplish this, we utilize a recently proposed generative model over protein structures, RFdiffusion \citep{rfdiff}. RFdiffusion unconditionally generates backbone structures of protein monomers. Successfully optimizing sequences that fold to generated structures has the additional benefit that it suggests a path forward for a full end-to-end top down protein design pipeline, where desirable structures are conditionally generated with RFdiffusion and then inverse folded with \texttt{BO-IF}.

We generate 24 protein backbones using RFdiffusion, each ranging between 100 to 150 amino acids in length. We perform focused evaluations on particular structures, evaluating a total of 150,000 sequences for each backbone, requiring roughly a GPU day per structure for both methods. We believe these samples serve as a challenging and unbiased benchmark for assessing the performance of both conventional inverse folding methods and our optimization-focused approach.

As described before, we directly target the (computationally determined) structural similarity between inverse folded sequences and the target structure as an objective. To measure structural similarity, we computationally fold a sequence, and compute similarity to the target structure according to two metrics: (1) TM Score as computed by \texttt{TM-algin} \citep{zhang2005tm}, and (2) RMSD.

\begin{figure}
    \centering
    \includegraphics[width=\textwidth]{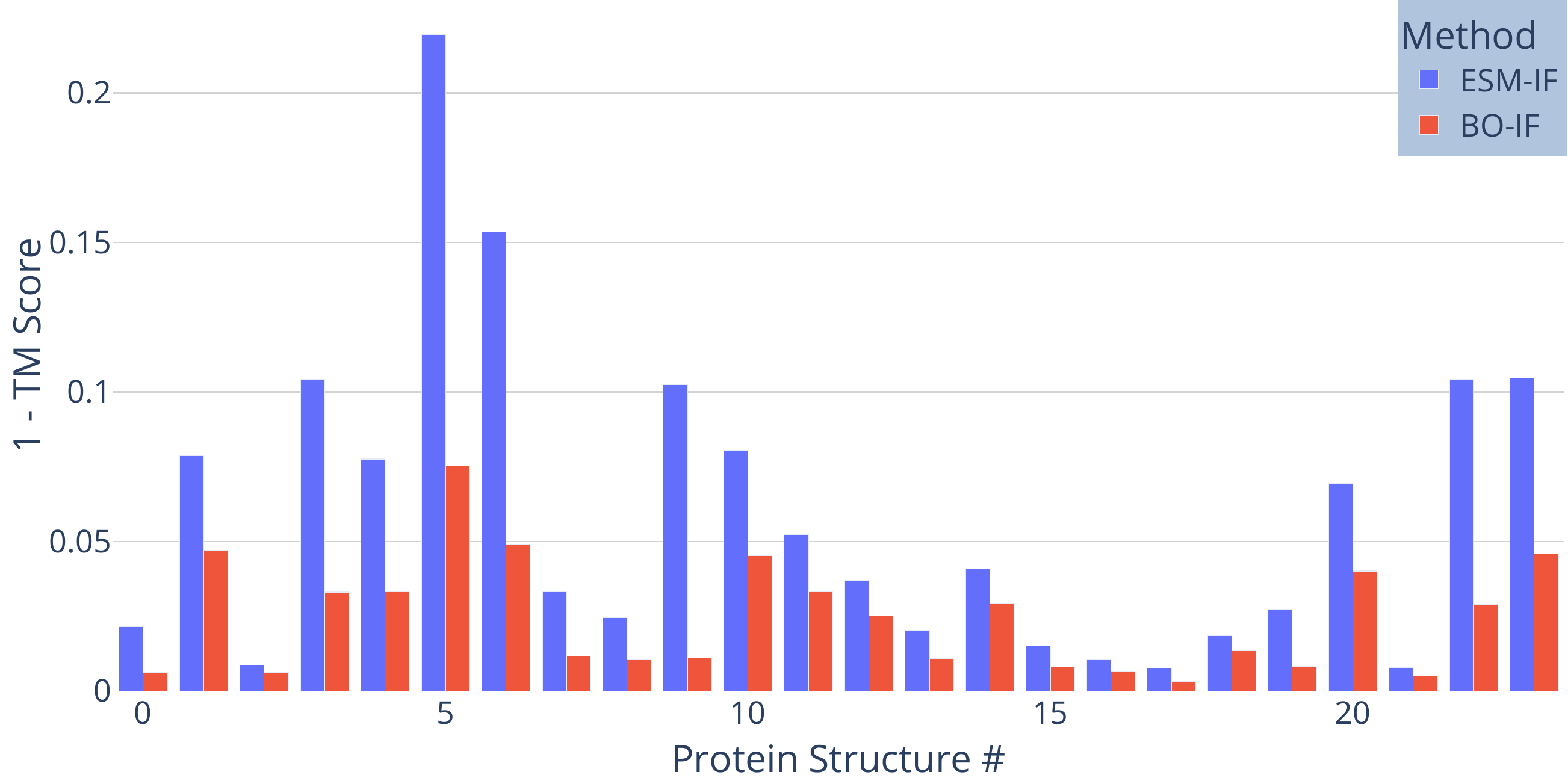}
    \caption{Error in backbone structures computationally folded from inverse folds as measured by \texttt{1-TM score} (computed by \texttt{TM-align}) across 24 target protein backbones. On average, optimization reduces structural error by \textbf{48\%} (standard error $\pm$ 0.69\%). This corresponds to RMSD reduction of \textbf{28\%} (standard error $\pm$ 0.81\%) on average.}
    \label{fig:tm_bar}
\raggedbottom
\end{figure}

\begin{figure}[ht]
\begin{center}
\includegraphics[width=\textwidth]{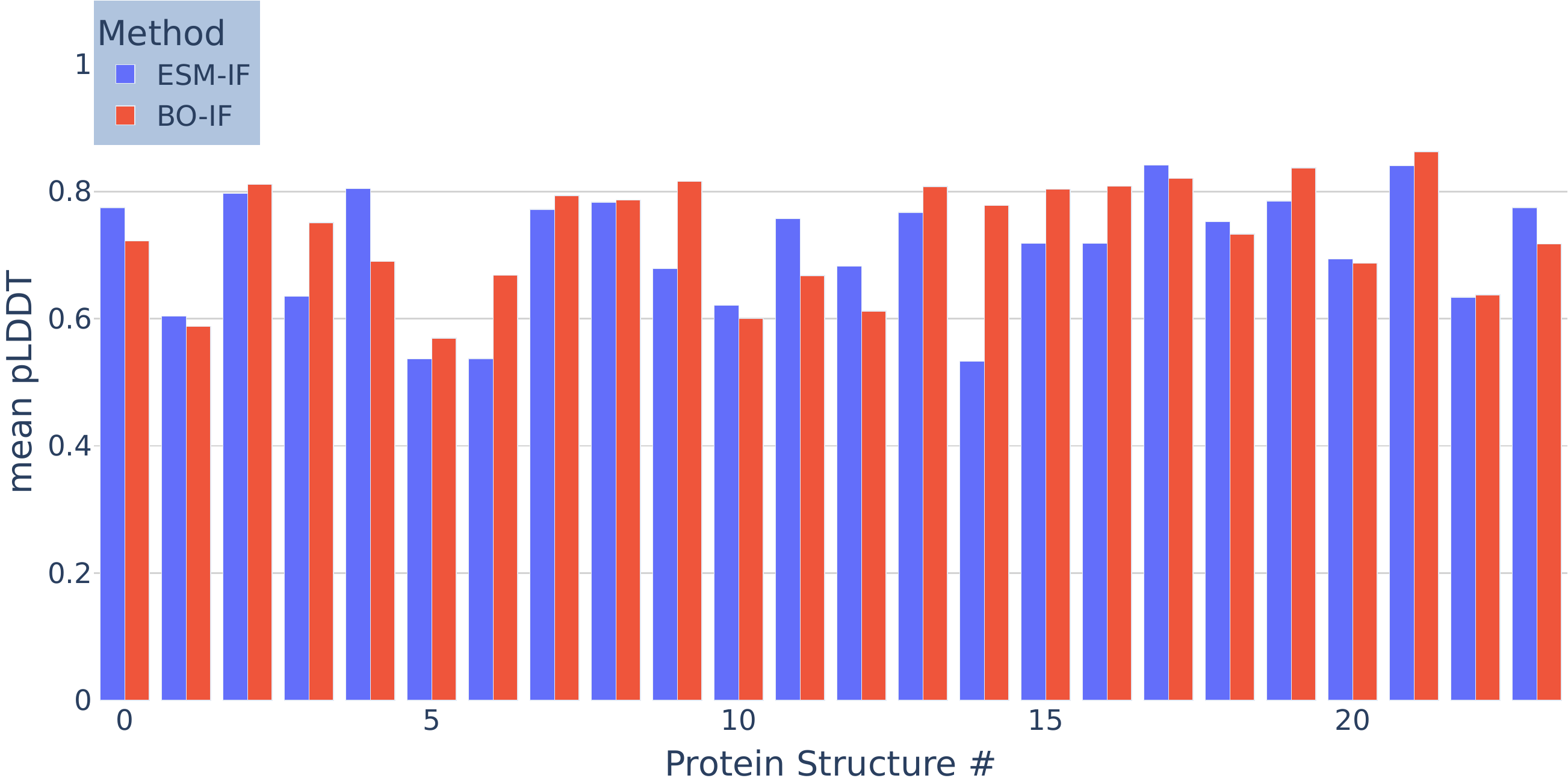} 
\hspace{1cm} 
\caption{Despite lower structural error, sequences produced by Bayesian Optimization (\texttt{BO-IF}) and {ESM-IF} display similar fold confidences on all protein backbones, with BO achieving a slightly higher mean pLDDT score of 0.732 compared to 0.710 with ESM-IF.}
\label{pLDDT_esmif_vs_bo}
\end{center}
\end{figure}

Results across all proteins are plotted in \autoref{fig:tm_bar}. Optimization produces sequences whose computationally determined structures have on average \textbf{48\%} lower structural error as measured by $1 - \textrm{TM Score}$, and never fails to improve TM Score across all structures. These correspond to a reduction in RMSD of \textbf{28\%} on average. In \autoref{pLDDT_esmif_vs_bo}, we find that ESM-IF and BO produce structures with comparable fold confidence as measured by pLDDT, with most proteins folding reasonably confidently on average. If higher scores are desired, this might be achievable, for example, by using pLDDT as a constraint.
\begin{table}[!htbp]
  \centering
  \caption{Details on model statistics and training}
  \label{}
  \small
  \begin{tabular}{cccccc}
    \hline
    \hline
    & Bayesian Optimization & GVP Transformer \\
    \hline
    Model Parameters & 47M & 142M \\
    GPUs & 1$\times$ RTX A6000 & 32$\times$ RTX 8000 \\
    Pretraining Time (GPU days) & 6 & 653 \\
    Optimization Runtime (GPU hours @ 150k evaluations) & 50 & 48 \\
    \hline
  \end{tabular} 
  \label{tab:vae_training}
\end{table}
\begin{figure}[ht] 
\begin{center}
\includegraphics[width=\columnwidth]{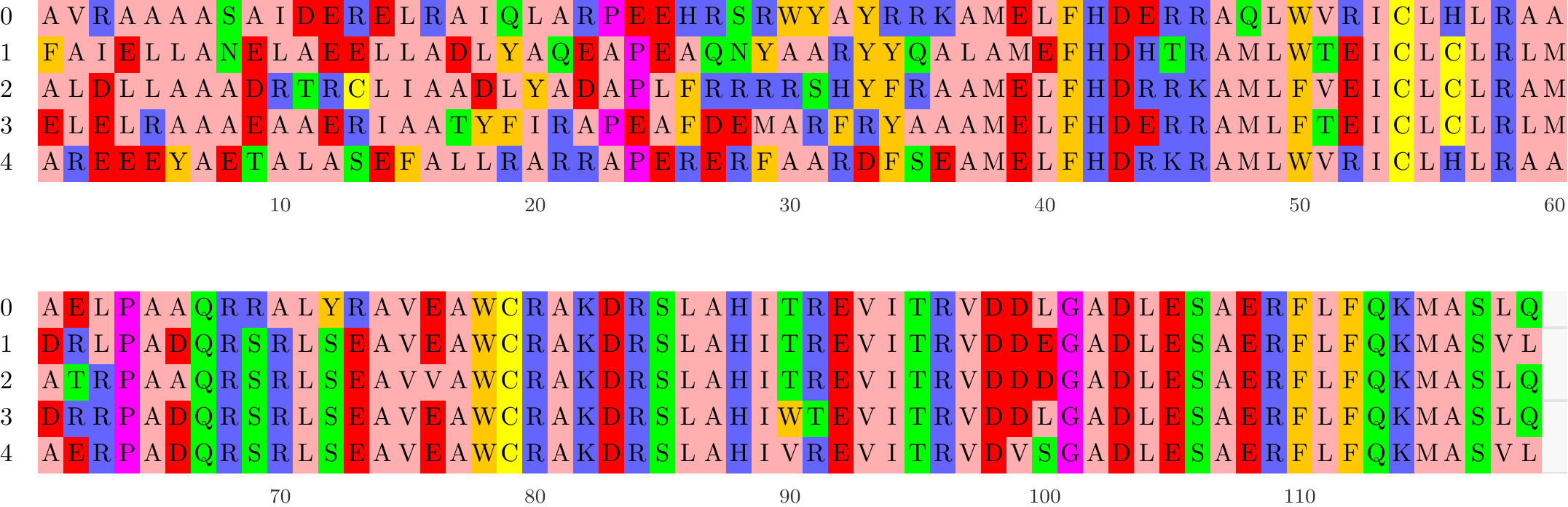} 
\hspace{1cm} 
\caption{Example multiple sequence alignment produced by optimization finding 5 diverse inverse folds for an input structure achieving an average TM score of $0.95 \pm 0.002$. 
Colors represent diversity in physicochemical properties. }
\label{fig:msa}
\end{center}
\vskip -0.2in
\raggedbottom
\end{figure}
\subsection{Optimization with constraints}
In this section, we present a case study on leveraging constrained Bayesian optimization. Protein engineering is often subject to diverse sequence constraints---discrete and continuous---to enable effective production, testing and validation of designed proteins in target applications. As an example constraint, immunogenicity describes the liability of a therapeutic molecule to produce an unwanted immune response in a patient, resulting in rapid drug clearance, adverse events and limitations of the drug's effectiveness. The humanness of a sequence---how similar it is to sequences found in the human proteome---is inversely correlated with the likelihood of immunogenicity, and thus a common constraint used in drug development.

\begin{figure}
    \centering
    \includegraphics[width=\textwidth]{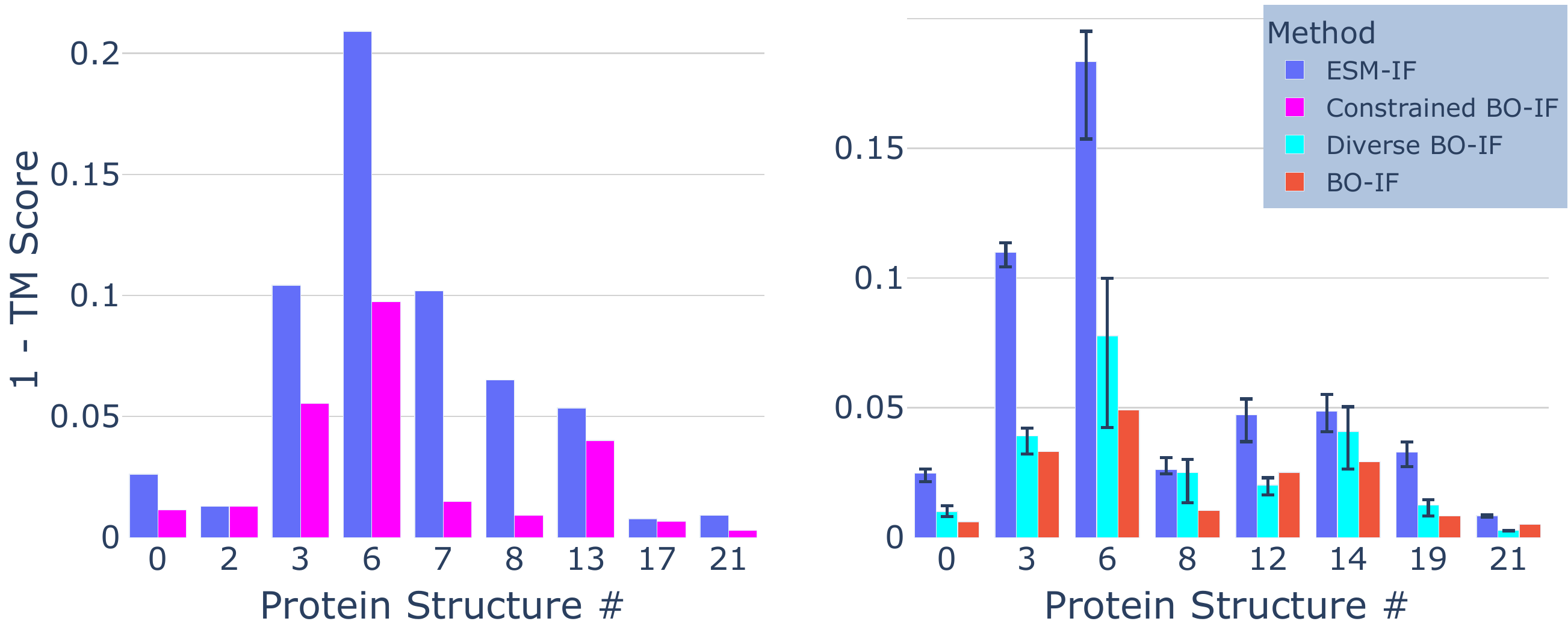}
    \caption{\textbf{(Left)} Inverse folding performed subject to the constraint that humanness is at least 0.8 (see text). 
     \textbf{(Right)} Finding diverse sets of $5$ inverse folds with pairwise edit distance $\geq 20$. Bar heights denote average value for each structure, with error bands denoting min and max values. The best single result achieved by standard \texttt{BO-IF} is included for comparison. See the appendix for additional results finding larger diverse sets of $10$ inverse folds. Both plots display a random subset of our generated structures.
    }
    \label{fig:cbo_robot_bar}
\end{figure}

We design inverse folds for $9$ protein structures subject to the constraint that humanness of the resulting sequence is at least $80\%$. 
To measure humanness, we fine-tune a classifier from the 150M parameter ESM2~\cite{lin2022language} model using 47,000 human and 64,000 non-human protein sequences downloaded from Uniref \citep{uniref}. 
Our classifier achieves a test accuracy of 93\%. Across all structures, we note that only 0.2\% of sequences generated using ESM-IF satisfy this constraint while also achieving TM score > 0.8. 
In \autoref{fig:cbo_robot_bar} \textbf{(Left)}, we show results comparing sequence optimization under this constraint to sequences generated from ESM-IF. For generated sequences, we discard sequences that fail to meet the humanization threshold. On average, optimization reduces structural error by 48.364\% (standard error $\pm$ 9.480\%) and RMSD by 34.454\% (standard error $\pm$ 7.082\%) compared to ESM-IF.

\subsection{Finding Diverse Sequences}
We apply \robot{} \citep{maus2022discovering} to seek a diverse set of 5 sequences as described in \autoref{sec:extensions}. We define two sequences to be diverse if they have a minimum edit distance of 20 (i.e., $\tau = 20$). \autoref{fig:cbo_robot_bar} \textbf{(Right)} shows results comparing TM scores achieved by \robot{} and ESM-IF for multiple target backbones. For each backbone, bar height denotes the average $1-$TM score achieved across the 5 sequences found for that structure, and error bars represent the range of TM scores (i.e., min and max). For comparison, we also plot the best sequence found by BO. Taking the average across these mean TM-scores, optimization reduces structural error by 48.640\% (standard error $\pm$ 8.003\%) and RMSD by 33.617\% (standard error $\pm$ 5.520\%) compared to ESM-IF. In addition to summary statistics, we display an example multiple sequence alignment for the diverse inverse folds found in \autoref{fig:msa}. See the appendix for additional results seeking a diverse set of 10 sequences (rather than only 5) as well as multiple sequence alignments for all diverse sets of inverse folds optimized.

\section{Discussion and Limitations}
\label{sec:discussion}

Generative modeling for protein structures is an exciting new technology. However, translating generated structures into physical proteins demands a solution to inverse folding: knowing what a final product should look like is not the same as knowing how to make it. We have demonstrated that Bayesian optimization can be an effective strategy for focused inverse folding of particular structures.

\textbf{Limitations.} While there is clearly promise in optimization for focused inverse folding, we clarify a few important limitations of our method. First, we are considering on the order of tens or hundreds of thousands of sequences per structure, rather than one shot predictions. While our approach yields substantial improvements for particular target structures, considering hundreds of thousands of sequences for each of a large dataset of proteins isn't feasible.

Perhaps most crucially, the objective functions we consider in this paper depend directly on computational folding. If Alphafold or ESMFold are inaccurate, so are our structures. This is notably in contrast to other metrics like native sequence recovery, where by definition recovering the ``known'' sequence achieves the correct fold. That is not to say that native sequence recovery is not without its own problems: notably, it can only be used for structures for which native sequences are available (i.e., not generated structures), and as a metric it discourages sequence diversity. 

Nevertheless, our results clearly demonstrate that inverse folding through Bayesian optimization is a compelling and efficient potential alternative approach to one-shot prediction in situations where the goal is to develop sequences that achieve the best possible map to an individual target structure. Furthermore, leveraging the optimization literature brings with it considerable advantages, such as the ability to handle constraints, multiple objectives, etc.


\small
\printbibliography
\clearpage

\clearpage
\appendix

\noindent\rule{\textwidth}{1pt}
\begin{center}
\vspace{7pt}
{\Large  Appendix}
\end{center}
\noindent\rule{\textwidth}{1pt}
\label{sec:appendix}

\section{Additional results for diverse Bayesian optimization}
\label{sec:results2} 
In this section we provide additional diverse optimization results. 

\begin{figure}[h]
    \centering
    \includegraphics[width=\textwidth]{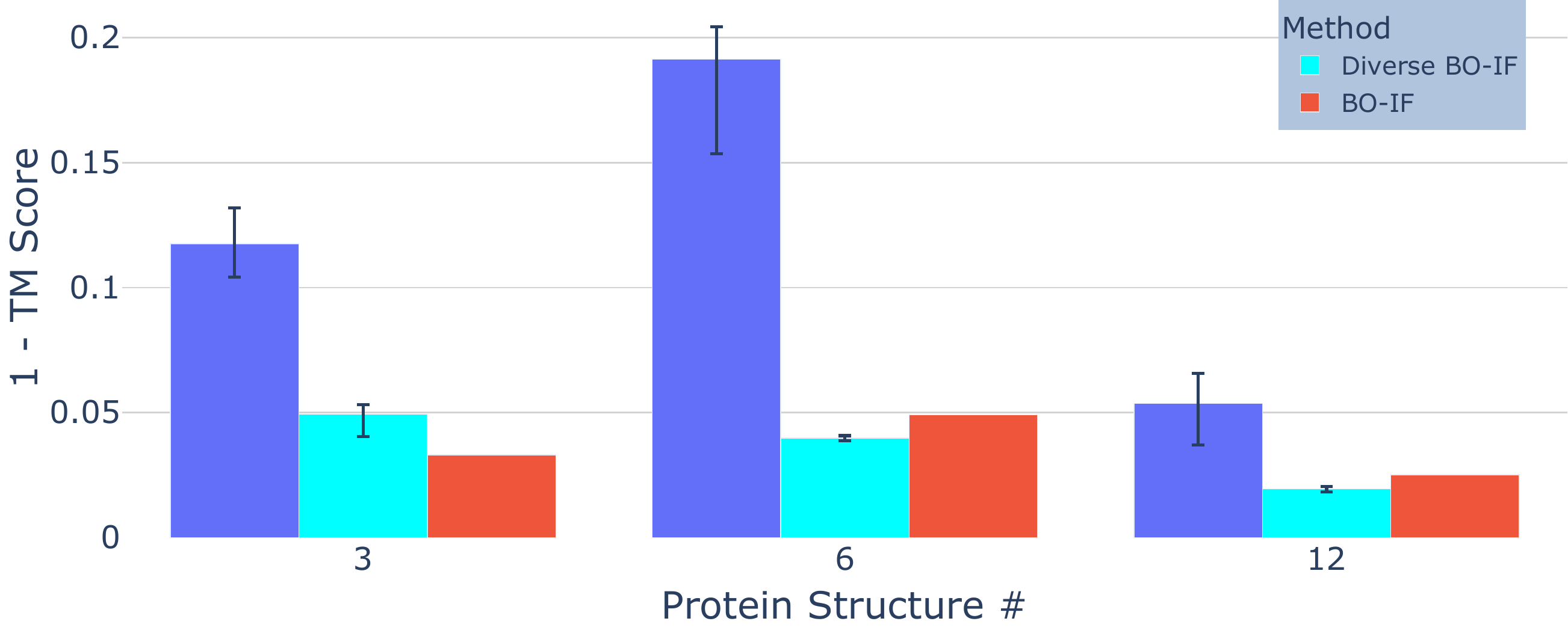}
    \caption{Finding 10 diverse inverse folds with a pairwise edit distance $\geq 5$. Bar heights denote average value of the 10 sequences found for each structure, with error bands denoting min and max values. The best single result achieved by standard \texttt{BO-IF} is included for comparison. Plot displays a random subset of 3 of our generated structures.
    }
    \label{fig:robot10}
\end{figure}

In \autoref{fig:cbo_robot_bar} \textbf{(Right)}, we provide results for applying \robot{} \citep{maus2022discovering} to seek a diverse set of $5$ inverse folds for $8$ randomly selected protein structures, where two sequences are defined to be diverse if they have a minimum edit distance of $20$.
In \autoref{fig:robot10}, we provide similar results for seeking a larger diverse set of inverse folds (diverse sets of $10$ inverse folds) for $3$ randomly selected structures. Here since we seek a larger diverse set, we use a looser diversity constraint, defining two sequences to be diverse if they have a minimum edit distance of $5$. 

\begin{figure}[h]
    \centering
    \includegraphics[width=\textwidth]{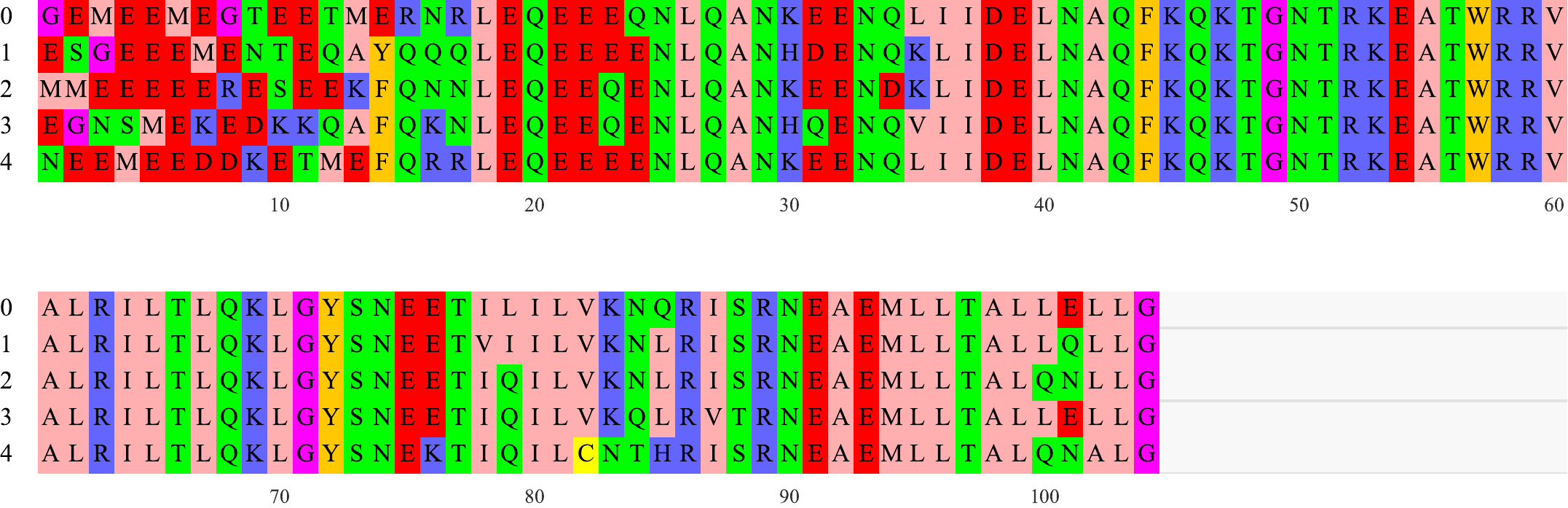}
    \caption{Multiple sequence alignment produced by optimization finding $5$ diverse inverse folds for the protein structure number $0$. Colors represent diversity in physicochemical properties.}
    \label{fig:msa0}
\end{figure}

\begin{figure}[h]
    \centering
    \includegraphics[width=\textwidth]{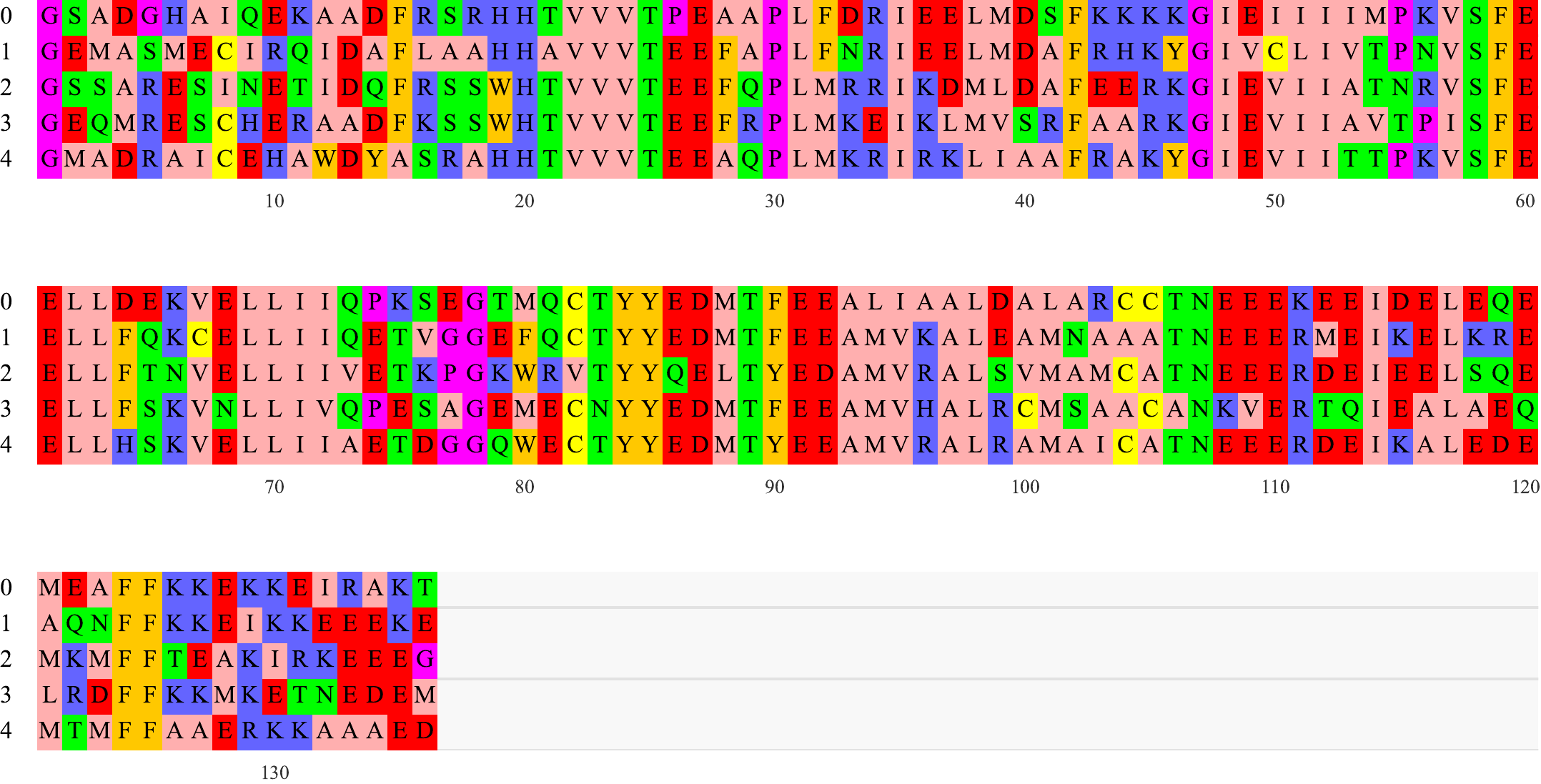}
    \caption{Multiple sequence alignment produced by optimization finding $5$ diverse inverse folds for the protein structure number $3$. Colors represent diversity in physicochemical properties.}
    \label{fig:msa3}
\end{figure}

\begin{figure}[h]
    \centering
    \includegraphics[width=\textwidth]{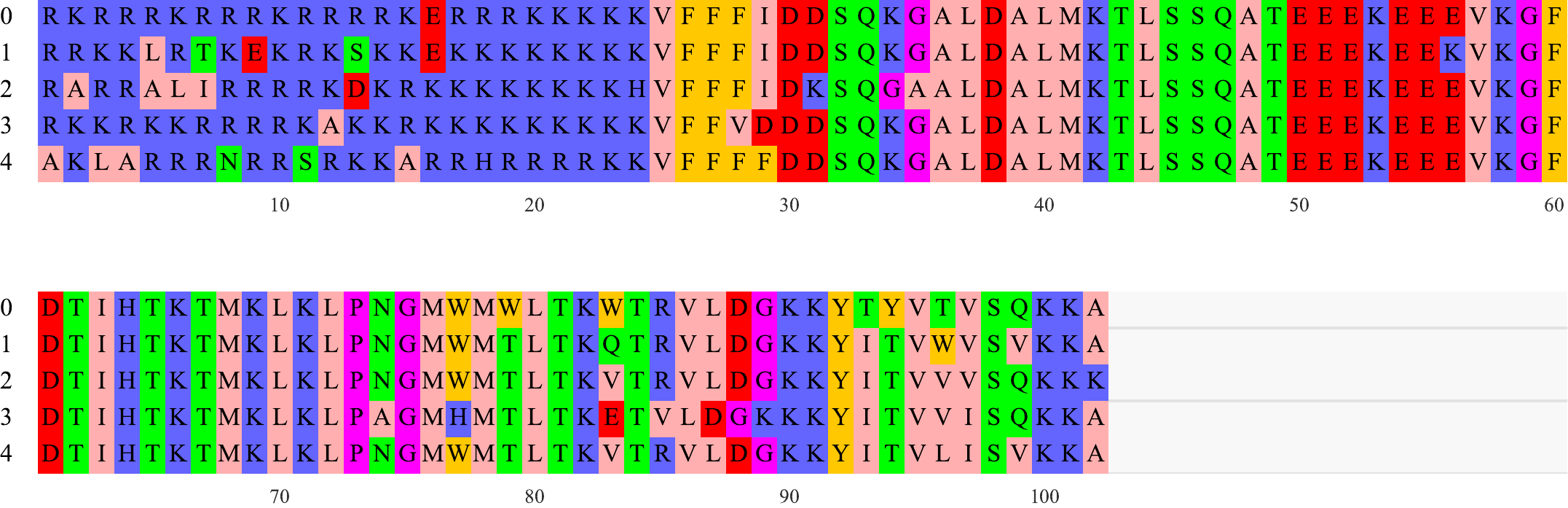}
    \caption{Multiple sequence alignment produced by optimization finding $5$ diverse inverse folds for the protein structure number $6$. Colors represent diversity in physicochemical properties.}
    \label{fig:msa6}
\end{figure}

\begin{figure}[h]
    \centering
    \includegraphics[width=\textwidth]{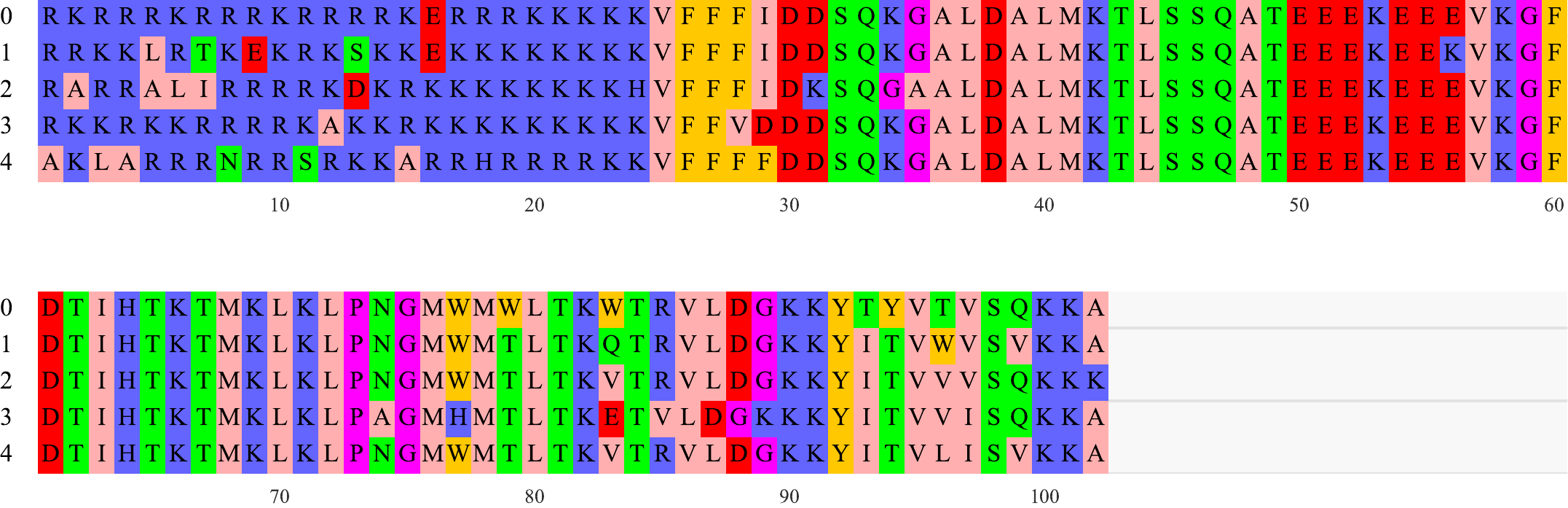}
    \caption{Multiple sequence alignment produced by optimization finding $5$ diverse inverse folds for the protein structure number $8$. Colors represent diversity in physicochemical properties.}
    \label{fig:msa8}
\end{figure}

\begin{figure}[h]
    \centering
    \includegraphics[width=\textwidth]{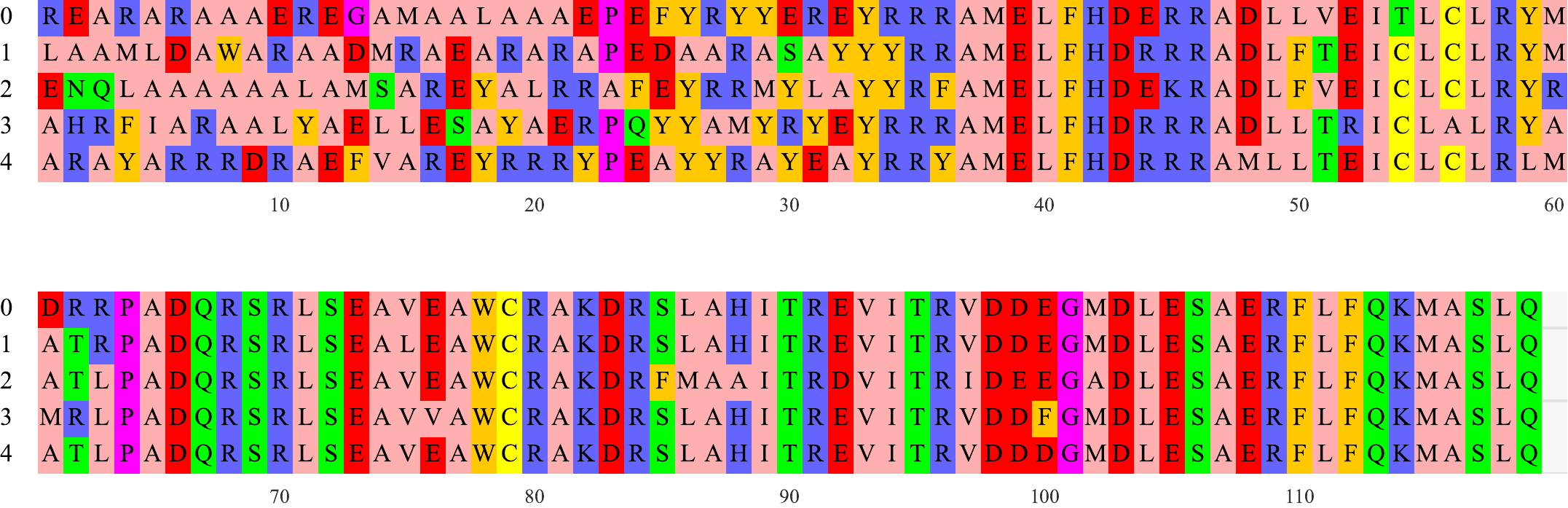}
    \caption{Multiple sequence alignment produced by optimization finding $5$ diverse inverse folds for the protein structure number $12$. Colors represent diversity in physicochemical properties.}
    \label{fig:msa12}
\end{figure}

\begin{figure}[h]
    \centering
    \includegraphics[width=\textwidth]{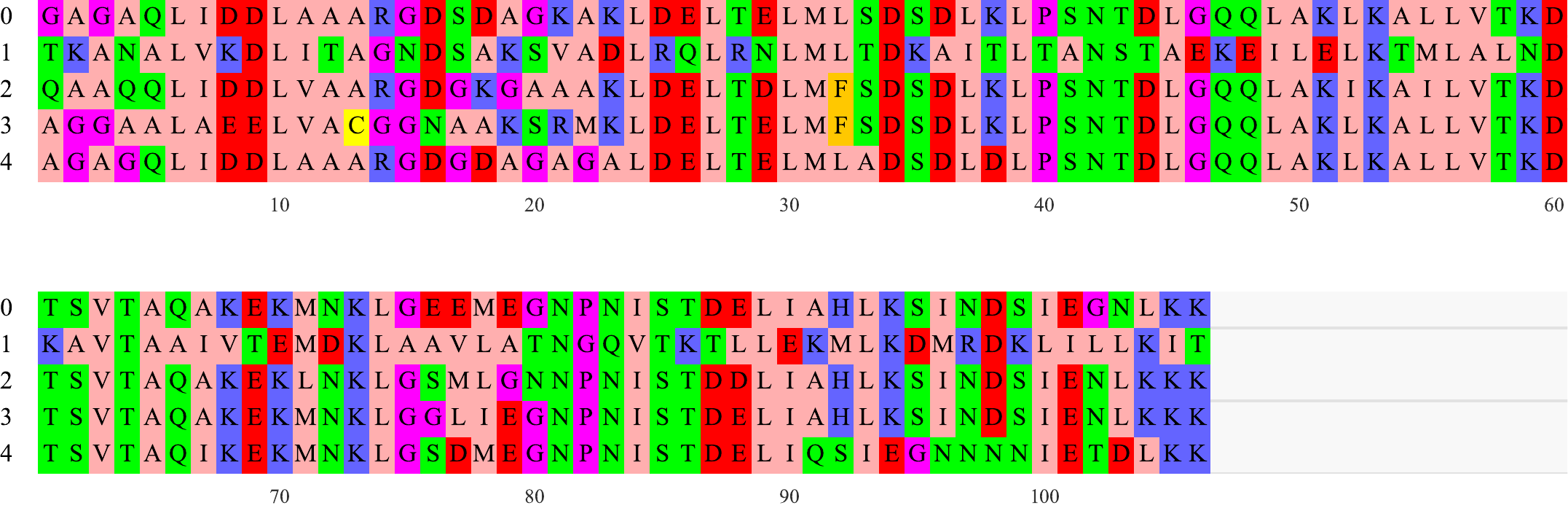}
    \caption{Multiple sequence alignment produced by optimization finding $5$ diverse inverse folds for the protein structure number $14$. Colors represent diversity in physicochemical properties.}
    \label{fig:msa14}
\end{figure}

\begin{figure}[h]
    \centering
    \includegraphics[width=\textwidth]{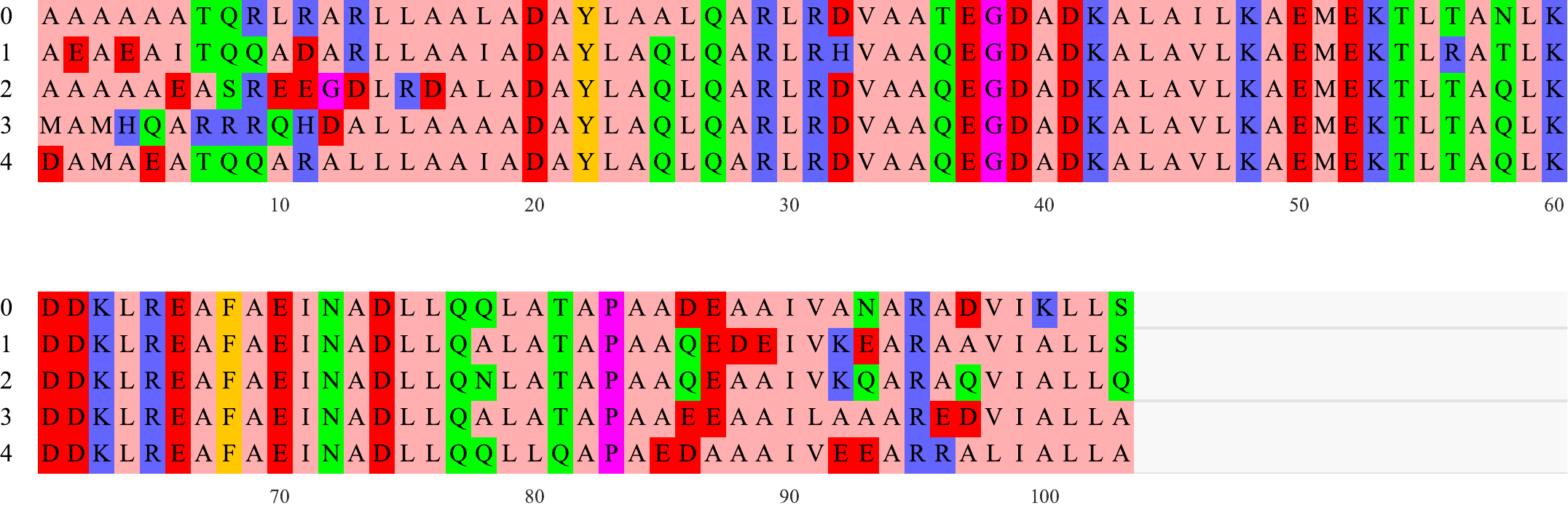}
    \caption{Multiple sequence alignment produced by optimization finding $5$ diverse inverse folds for the protein structure number $19$. Colors represent diversity in physicochemical properties.}
    \label{fig:msa19}
\end{figure}

\begin{figure}[h]
    \centering
    \includegraphics[width=\textwidth]{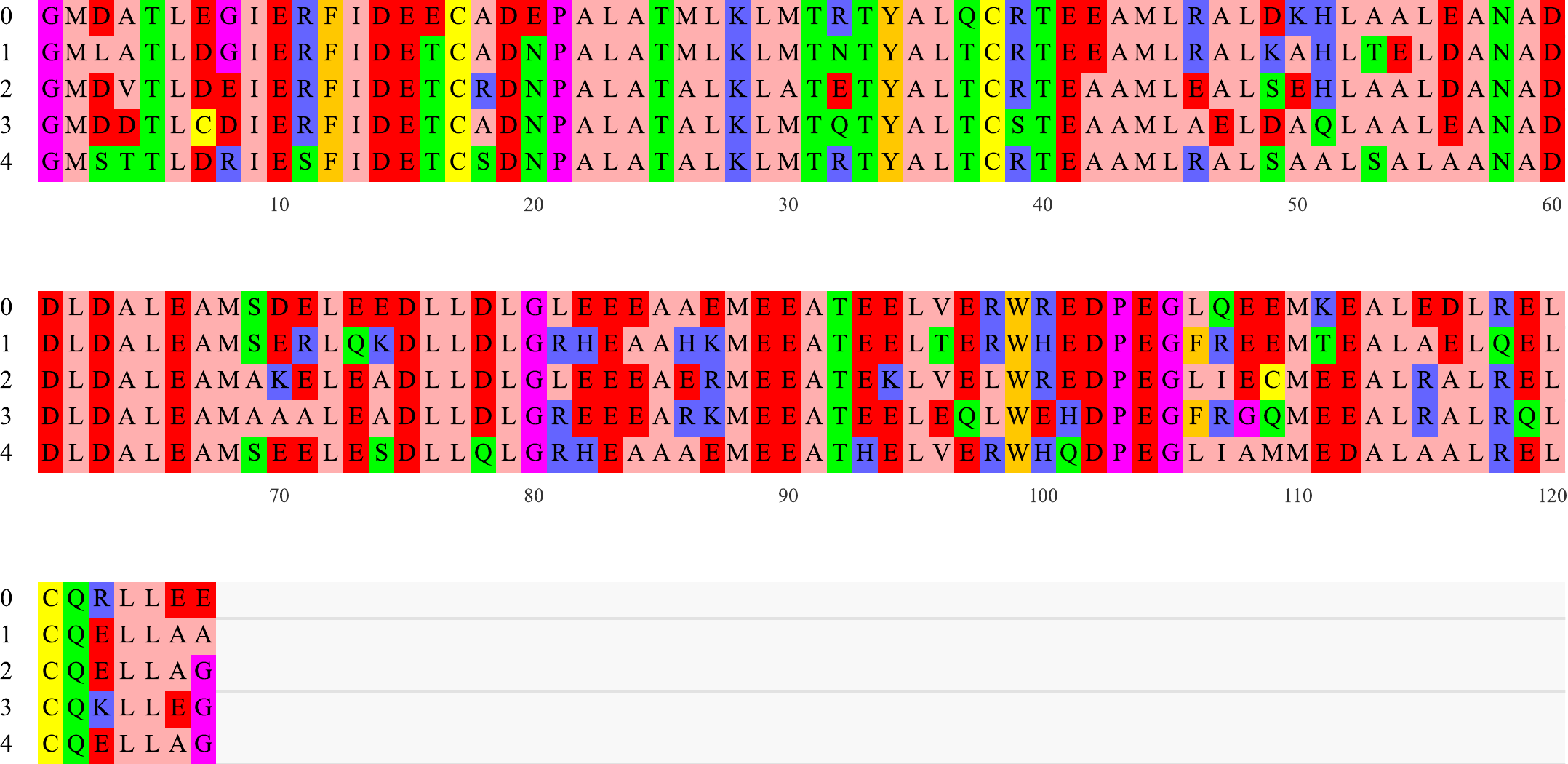}
    \caption{Multiple sequence alignment produced by optimization finding $5$ diverse inverse folds for the protein structure number $21$. Colors represent diversity in physicochemical properties.}
    \label{fig:msa21}
\end{figure}

\begin{figure}[h]
    \centering
    \includegraphics[width=\textwidth]{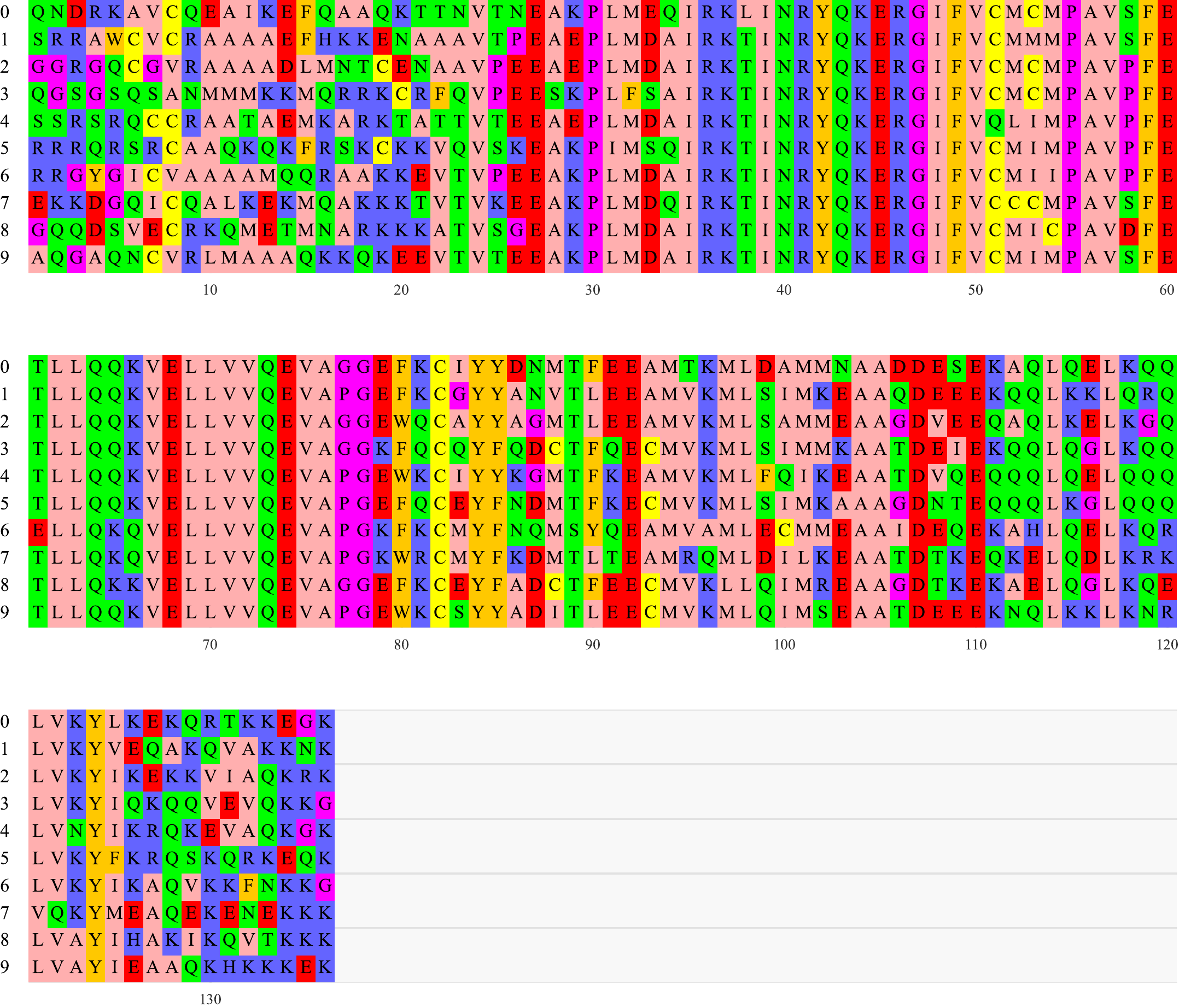}
    \caption{Multiple sequence alignment produced by optimization finding $10$ diverse inverse folds for the protein structure number $3$. Colors represent diversity in physicochemical properties.}
    \label{fig:msa3_10}
\end{figure}

\begin{figure}[h]
    \centering
    \includegraphics[width=\textwidth]{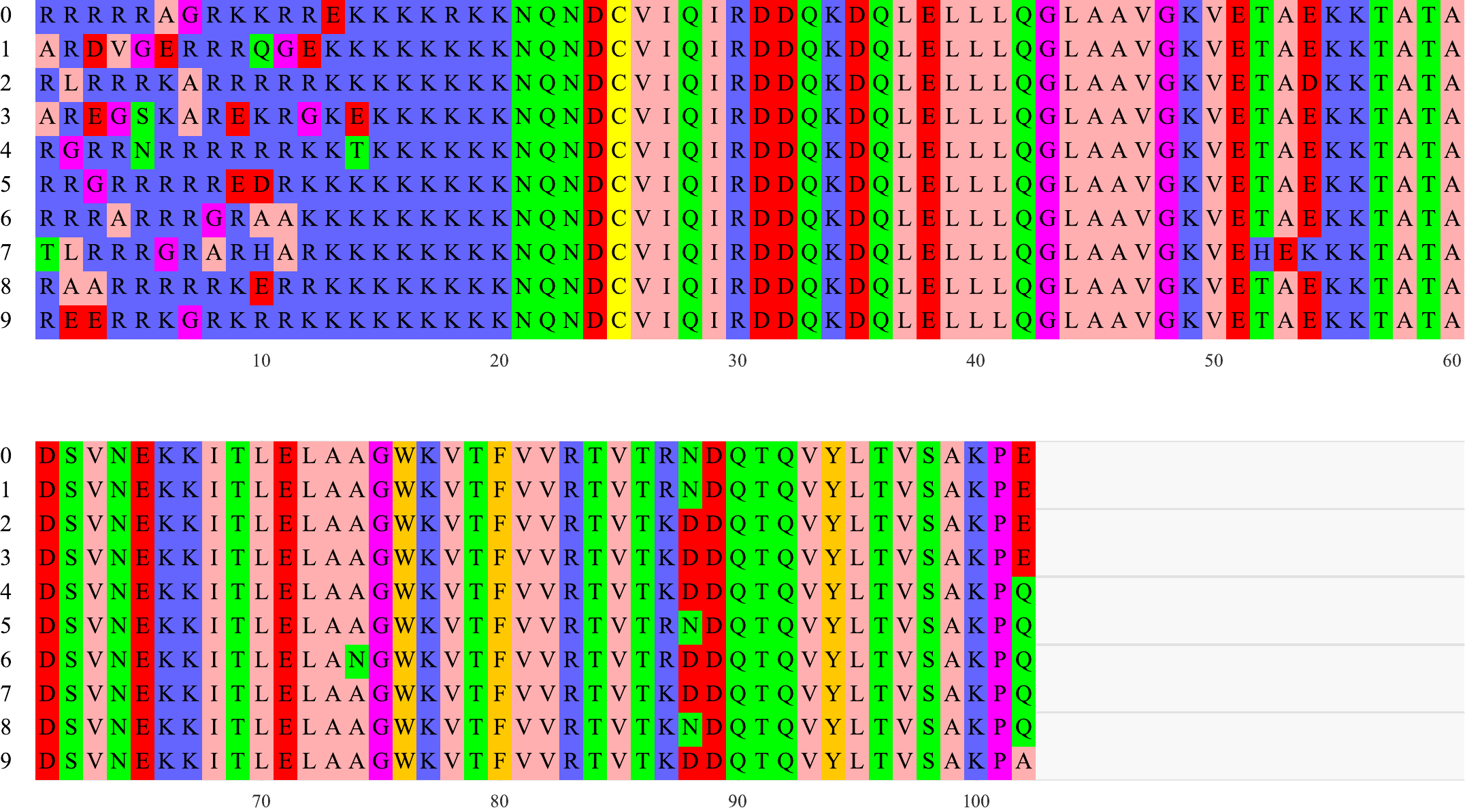}
    \caption{Multiple sequence alignment produced by optimization finding $10$ diverse inverse folds for the protein structure number $6$. Colors represent diversity in physicochemical properties.}
    \label{fig:msa6_10}
\end{figure}

\begin{figure}[h]
    \centering
    \includegraphics[width=\textwidth]{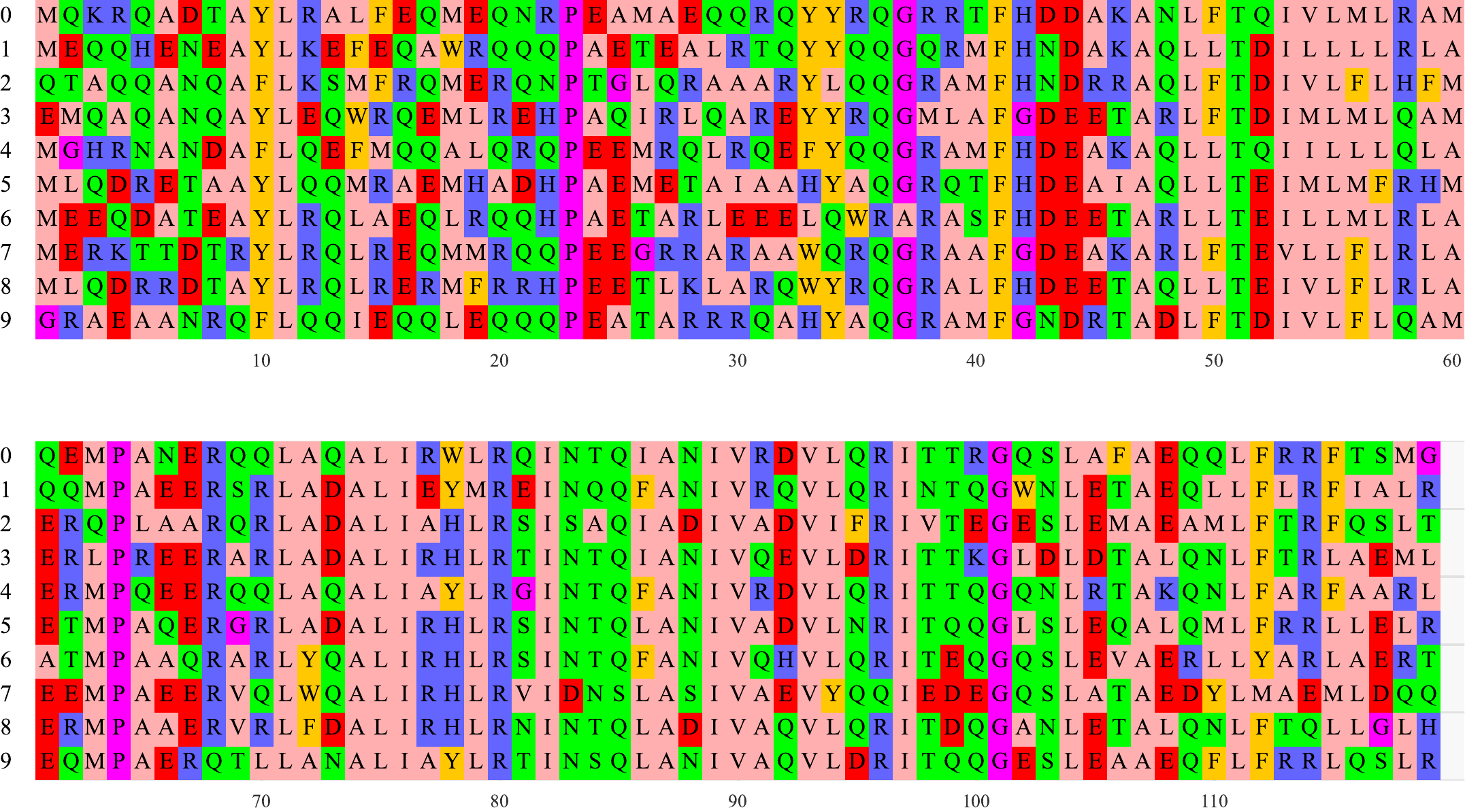}
    \caption{Multiple sequence alignment produced by optimization finding $10$ diverse inverse folds for the protein structure number $12$. Colors represent diversity in physicochemical properties.}
    \label{fig:msa12_10}
\end{figure}

Additionally, we display multiple sequence alignments for all diverse inverse folds in \autoref{fig:msa0} - \autoref{fig:msa12_10}. 

%


\end{document}